\documentclass{aastex61}

\shorttitle{Brightest LMXB in M31}
\shortauthors{Marelli et al.}

\begin{document}

\title{Discovery of periodic dips in the brightest hard X-ray source of M31 with EXTraS}

\correspondingauthor{Martino Marelli}
\email{marelli@lambrate.inaf.it}

\author{Martino Marelli}
\affiliation{Scuola Universitaria Superiore IUSS Pavia, piazza della Vittoria 15, 27100 Pavia, Italy}
\affiliation{INAF - Istituto di Astrofisica Spaziale e Fisica Cosmica Milano, via E. Bassini 15, 20133 Milano, Italy}

\author{Andrea Tiengo}
\affiliation{Scuola Universitaria Superiore IUSS Pavia, piazza della Vittoria 15, 27100 Pavia, Italy}
\affiliation{INAF - Istituto di Astrofisica Spaziale e Fisica Cosmica Milano, via E. Bassini 15, 20133 Milano, Italy}
\affiliation{INFN - Istituto Nazionale di Fisica Nucleare, Sezione di Pavia, Via Bassi 6, 27100 Pavia, Italy}

\author{Andrea De Luca}
\affiliation{INAF - Istituto di Astrofisica Spaziale e Fisica Cosmica Milano, via E. Bassini 15, 20133 Milano, Italy}
\affiliation{INFN - Istituto Nazionale di Fisica Nucleare, Sezione di Pavia, Via Bassi 6, 27100 Pavia, Italy}

\author{David Salvetti}
\affiliation{INAF - Istituto di Astrofisica Spaziale e Fisica Cosmica Milano, via E. Bassini 15, 20133 Milano, Italy}

\author{Luca Saronni}
\affiliation{Universit\'a degli Studi di Milano, via Festa del Perdono 7, 20122 Milano, Italy}

\author{Lara Sidoli}
\affiliation{INAF - Istituto di Astrofisica Spaziale e Fisica Cosmica Milano, via E. Bassini 15, 20133 Milano, Italy}

\author{Adamantia Paizis}
\affiliation{INAF - Istituto di Astrofisica Spaziale e Fisica Cosmica Milano, via E. Bassini 15, 20133 Milano, Italy}

\author{Ruben Salvaterra}
\affiliation{INAF - Istituto di Astrofisica Spaziale e Fisica Cosmica Milano, via E. Bassini 15, 20133 Milano, Italy}

\author{Andrea Belfiore}
\affiliation{Scuola Universitaria Superiore IUSS Pavia, piazza della Vittoria 15, 27100 Pavia, Italy}
\affiliation{INAF - Istituto di Astrofisica Spaziale e Fisica Cosmica Milano, via E. Bassini 15, 20133 Milano, Italy}

\author{Gianluca Israel}
\affiliation{INAF -- Osservatorio Astronomico di Roma, via Frascati 33, 00040 Monteporzio Catone, Italy}

\author{Frank Haberl}
\affiliation{Max-Planck-Institut fur extraterrestrische Physik, Giessenbachstrase, 85748 Garching, Germany}

\author{Daniele D'Agostino}
\affiliation{Istituto di Matematica Applicata e Tecnologie Informatiche CNR, via dei Marini 6, 16149 Genova, Italy}

\begin{abstract}

We performed a search for eclipsing and dipping sources in the archive of the EXTraS project -- a systematic characterization of the temporal
behaviour of {\it XMM-Newton} point sources.
We discovered dips in the X-ray light curve of 3XMM J004232.1+411314, which has been recently associated with the hard X-ray source dominating the emission of M31.
A systematic analysis of {\it XMM-Newton} observations revealed 13 dips in 40 observations (total exposure time $\sim$0.8 Ms).
Among them, four observations show two dips, separated by $\sim$4.01 hr. Dip depths and durations are variable.
The dips occur only during low-luminosity states (L$_{0.2-12}<1\times10^{38}$ erg s$^{-1}$),
while the source reaches L$_{0.2-12}\sim2.8\times10^{38}$ erg s$^{-1}$.
We propose this system to be a new dipping Low-Mass X-ray Binary in M31 seen at high inclination (60$^{\circ}$--80$^{\circ}$),
the observed dipping periodicity is the orbital period of the system.
A blue HST source within the {\it Chandra} error circle is the most likely optical counterpart of the accretion disk.
The high luminosity of the system makes it the most luminous dipper known to date.
  
\end{abstract}

\keywords{stars: neutron --- pulsars: general --- galaxies: individual (M31) --- galaxies: bulges --- X-rays: binaries}

\section{Introduction} \label{sec:intro}

The EXTraS project (Exploring the X-ray Transient and variable Sky, \citet{del16}) developed
new techniques and tools to extract and describe the timing behavior of X-ray sources. The entire public {\it XMM-Newton} archive
(3XMM catalog, data release 4\footnote{\url{http://xmmssc-www.star.le.ac.uk/Catalogue/3XMM-DR4}}) was analyzed, obtaining hundreds of parameters to describe
the periodic and aperiodic variability of more than 500,000 serendipitous sources, on different time scales (from seconds to years). The results are
publicly available, together with a detailed documentation\footnote{\url{http://www.extras-fp7.eu}}.
Taking advantage from this improved {\it XMM-Newton} timing analysis, we performed a systematic search for eclipsing and dipping objects,
finding significant dips in the light curves of some observations of 3XMM J004232.1+411314 (XMM0042 hereafter).

XMM0042 is a moderately bright ($\sim$0.2 counts s$^{-1}$) source observed multiple times by {\it XMM-Newton} due to its proximity to the M31 bulge.
Based on flux and spectral variability studies in the 0.5-10 keV energy band,
the source has been classified as an X-ray binary candidate in various {\it Chandra} and {\it XMM-Newton} studies \citep[see e.g.][]{kon02,sti11}.
Based on {\it NuSTAR} data, it has been recently associated with the most prominent source of hard X-rays from M31 \citep{yuk17}.
The simultaneous {\it Swift/XRT} and {\it NuSTAR} spectrum is well fitted assuming an accretion disk model with a temperature of $\sim$0.2 keV and a broken power law,
with a photon index of $\sim$1 and an energy cutoff at $\sim$18 keV.
If XMM0042 is in M31 \citep[784$\pm$13 kpc,][]{sta98}, its 0.5-50 keV luminosity is $L_{0.5-50keV}\sim4\times10^{38}$ erg s$^{-1}$.
The most precise X-ray position of XMM0042 comes from {\it Chandra}, at R.A. (J2000) 00$^h$42$^m$32$^s$.072, Dec (J2000) +41$^{\circ}$13$'$14$''$.33 (0.4$''$, 3$\sigma$ error) \citep{bar14}.
{\it HST} observations revealed 17 possible optical/UV counterparts \citep{yuk17}, none of which is compatible with a high-mass donor ($>3M_{\odot}$).

We collected all the EXTraS results of XMM0042 and applied the same analysis to extract the same products from the most recent observations within the 3XMM catalog,
data release 7\footnote{\url{http://xmmssc.irap.omp.eu/Catalogue/3XMM-DR7/3XMM_DR7.html}}.
Section \ref{sec:analysis} describes the spectral study we performed, as well as the investigation of the source light curve.
The intepretation and discussion of our most relevant results are reported in Section \ref{sec:disc}, together with a discussion of the nature of the source in the light of these new results.

\section{Data analysis} \label{sec:analysis}

We searched for all the {\it XMM-Newton} observations of XMM0042 in the 3XMM catalog DR7.
EXTraS data make use of the same filters as the 3XMM catalogs - energy band (0.2-12 keV), pattern and flags.
We selected exposures with the most stable attitude, also excluding the ones with the source partially outside the field of view or on CCD gaps.
The observations we analyzed are listed in Appendix (Table \ref{tab:lc3a}).

We made use of SAS v.15 to perform a standard analysis from ODF files.
For the spectral analysis only, we excluded very-high-background periods ($>$40 counts/s from the PN camera and $>$15 from MOS1/2 from the entire field of view, in the 0.2-12 keV energy range).
Following \citet{yuk17} we adopted an absorbed \citep[abundances from][]{wil00} accretion disk model plus power-law (the cutoff energy is $\sim$17 keV, above our energy range).
As in \citet{yuk17}, the column density was fixed to the Galactic value of 7$\times10^{20}$ cm$^{-2}$.
Adding together the spectra of all the observations (and correcting for response matrices and effective areas), we obtain a poor fit, with a null hypothesis probability
nhp=1.5$\times10^{-10}$, 2312 degrees of freedom (dof).
The residuals are structured, with a clear shortage around 0.6 keV and an excess at around 1 keV.
The spectrum is well fitted either by adding a broad Gaussian emission line at $\sim$0.95 keV (nhp=9.5$\times10^{-3}$, dof=2310) or a broad Gaussian absorption line at
$\sim$0.6 keV (nhp=4.0$\times10^{-2}$, dof=2310). The best fitting parameters are reported in Table \ref{tab:spec}. A contemporaneous fit of all spectra is in agreement
with the single-spectra result.

\begin{figure}
\plotone{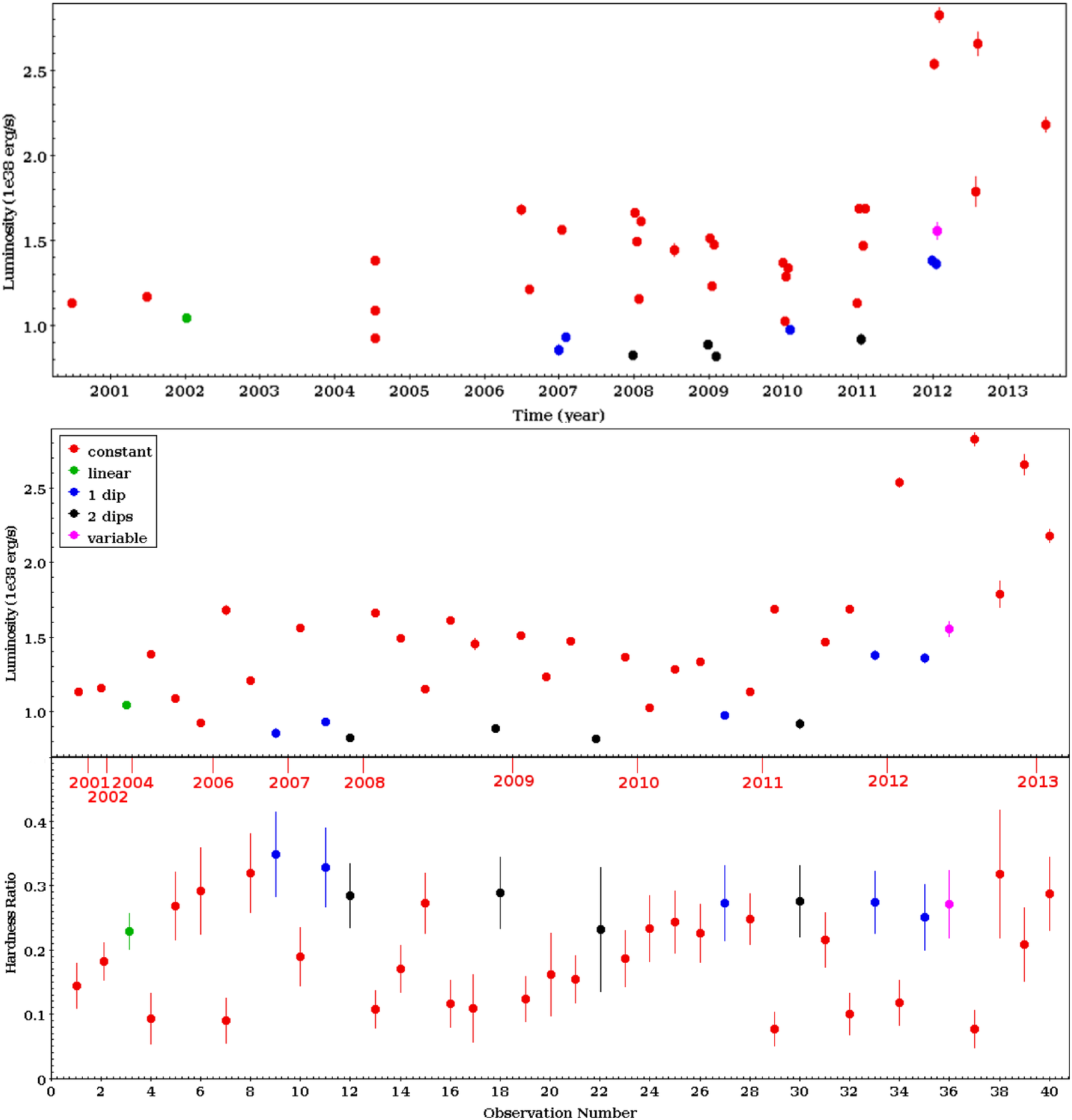}
\caption{{\it Upper panel:} Persistent X-ray luminosities during the analyzed {\it XMM-Newton} observations, as reported in Table \ref{tab:lc}.
  Different colours mark observations with light curve
  best fitted by a different model: red for constant, green for linear, magenta for variable, blue for single dip, black for double dips.\\
  {\it Central panel:} We show the same luminosity as in Upper Panel. The observation number in the X-axis allows to easily associate the observation
  with the results reported in Appendix (Table \ref{tab:lc3b}). Different colours mark observations with light curve best fitted by a different model, as above.\\
  {\it Lower panel:} Hardness ratio of XMM0042 comparing 0.2-0.8 keV and 0.8-2 keV energy ranges. The hardness ratio is defined as $(L_{08-2}-L_{02-08})/(L_{08-2}+L_{02-08})$
  (see Section \ref{sec:analysis}). Different colours mark observations with light curve best fitted by a different model, as above.
  \label{fig:spec}}
\end{figure}

We adopted the total absorbed double-component plus emission line model and parameters in Table \ref{tab:spec} to derive the X-ray luminosity
for each instrument and observation and exploit the simultaneous observations of PN, MOS1 and MOS2.
This was used to obtain hardness ratios in different energy bands for each observation. Figure \ref{fig:spec} shows the hardness ratio between 0.2-0.8 keV and 0.8-2 keV bands.
The variation with time is apparent (nhp=1.6$\times10^{-11}$, dof=39).
Higher-energy band hardness ratio analysis revealed no significant change with time.
This suggests some type of variation with time in the thermal component (or in the lines, if present).

EXTraS light curves from different instruments and exposures are binned using the same time bins (a grid of 500-s-time-bins beginning with the zero {\it XMM-Newton} reference time),
therefore we calculated the weighted mean, bin by bin, of the luminosity curves from PN and MOS1/2 to obtain the total light curve of each observation.
Then, we fitted each observation light curve using a constant model. If the fit was not statistically acceptable (3$\sigma$), we used a more complex model.
We tried with a linear model and constant-plus-dips model. The last one has four parameters: $T_{min}$ is the time of minimum luminosity, $\Delta T$ is the duration of the dip, $L_{min}$
is the minimum luminosity and $L_{out}$ is the luminosity outside the dip. An f-test \citep{bev69} was used to confirm the statistical improvement by using the more complex models.\\
According to our investigation, 29 curves with exposures varying from $\sim$11 ks to $\sim$33 ks are constant, five curves reveal a single dip and four have two dips, for a total of 13 significant dips.
The longest observation (obs.id 0112570101), with a $\sim$64 ks exposure, is the only one that reveals a linear decrease in luminosity.
Only one curve, (obs.id 0674210501), is variable at $>4\sigma$ but does not fit with our classification due to the presence of a more complex variability.

\begin{figure}
\plotone{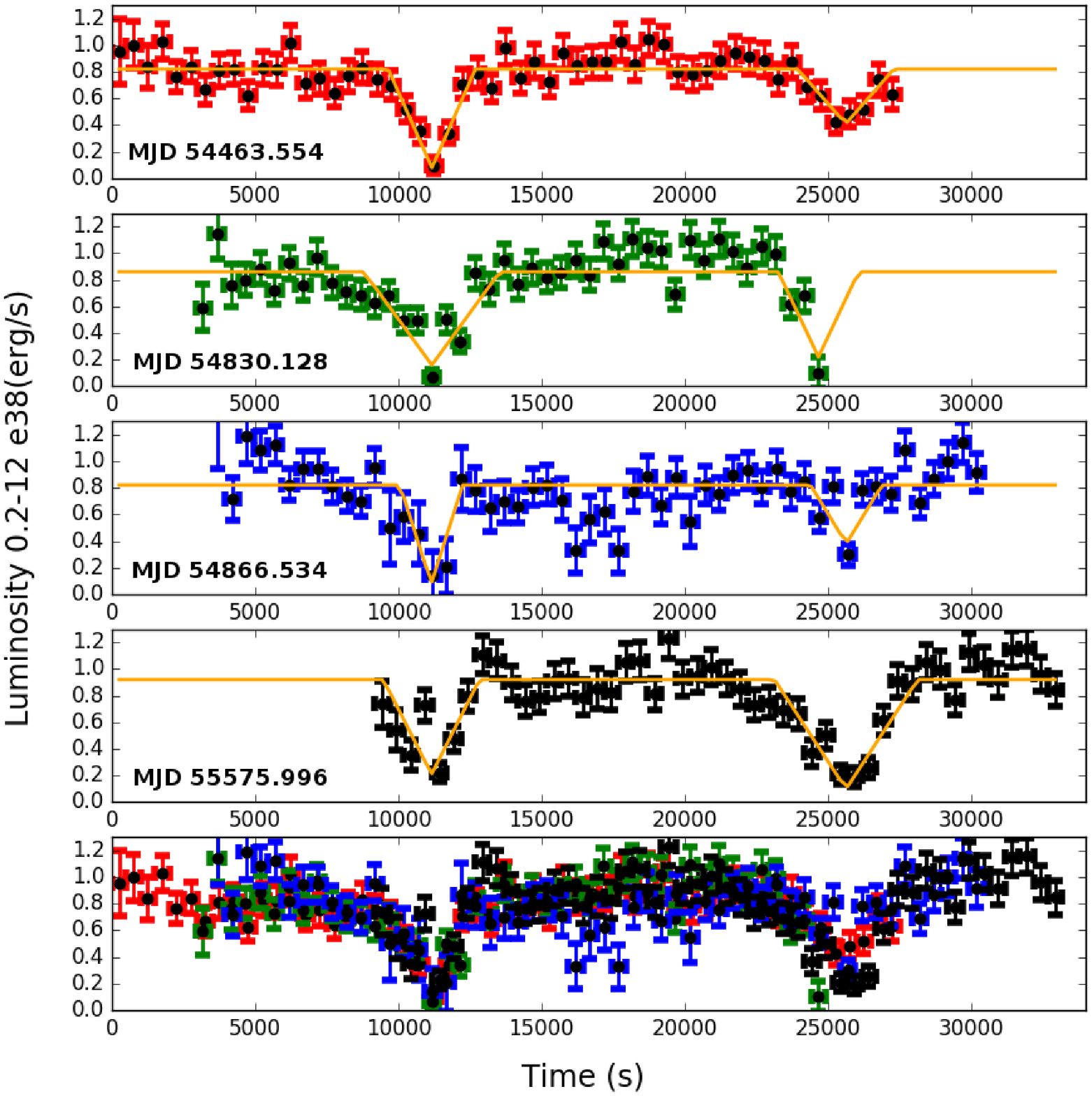}
\caption{The four {\it XMM-Newton} luminosity light curves (0.2-12 keV) of XMM0042 where two dips are found (observations 0505720201, 0551690201, 0551690601 and 0650560401 respectively).
   Each light curve start time is shown in the corresponding panel.
  They are aligned in order to have the first minimum of the model at the same time. We show the 1$\sigma$ error on 500-s time bins.
  The 4-hours period is apparent, as well as the dishomogenity of dip profiles. It is also apparent that outside the dips the source luminosity is consistent for
  all the observations. In the last panel, we show the superimposition of the curves.
  \label{fig:lc}}
\end{figure}

Within double-dipping observations, time interval between dips minima is consistent with being constant (1$\sigma$ confidence) revealing a periodicity of (14.47$\pm$0.12) ks,
 as apparent in Figure \ref{fig:lc}
(the minima separation in obs.id 0551690201 should be considered as a lower limit, because we observe only part of the second dip).
Considering a 14.47 ks period, all single-dip observations are characterized by exposures that do not allow for the detection of the previous and following dips.
On the other hand, almost all the observations well fitted by a constant model cover more than one period.\\
Searches for a periodicity using multiple data sets e.g. with the Lomb-Scargle algorithm \citep{zec09}
is hampered by the time separation between double-dipping observations.

We also divided the observations with dips in order to isolate the dipping periods from the rest of the observation.
We do not detect any significant spectral variation during the dips: a simulteneous fit of spectra of the two data sets,
with all the variables chained but a multiplicative factor, results in an acceptable fit (nhp=3.0$\times10^{-2}$, dof=1124).
We note that due to the low statistics we would not detect (3$\sigma$ confidence) variations in the disk temperature and in the photon index smaller than 25\% and 10\%, respectively.

The dip duration is variable from $\sim$2.2 ks to $\sim$5.0 ks; the minimum luminosity varies from $\sim$10\% to $\sim$70\% of the persistent luminosity.
We note that the low statistics prevents us to detect dips (3$\sigma$ significance) with a minimum luminosity $>$75\% of the constant luminosity.
Dipping observations occur only during low-luminosity states (L$_{0.2-12}<1\times10^{38}$ erg s$^{-1}$) while observations where we found no dips occur at all possible luminosities
($0.9\times10^{38}<$L$_{0.2-12}<1.7\times10^{38}$ erg s$^{-1}$) before 2012. Starting from 2012, the source persistent luminosity increases by a factor $\sim$2, with luminosities
$1.4\times10^{38}<$L$_{0.2-12}<2.8\times10^{38}$ erg s$^{-1}$. Dipping observations still occur only during low-luminosity states (see Figure \ref{fig:spec}).\\
The better spatial resolution of {\it Chandra} allows to detect two different sources, about 8$''$ apart, that contribute to the emission of the {\it XMM-Newton} XMM0042 source.
\citet{hof13} present an accurate study of the variability of these sources. Source 75 (R.A. (J2000) 00$^h$42$^m$32$^s$.07, Dec (J2000) +41$^{\circ}$13$'$14$''$.6) is the brightest one and varies
by a factor $\sim$5 during the 14 years of observation (from 1998 to 2012), showing a general increase of the flux with time on years-timescale.
Source 78 (R.A. (J2000) 00$^h$42$^m$32$^s$.74, Dec (J2000) +41$^{\circ}$13$'$11$''$.1) is almost constant, a factor 10 less luminous than source 75.
Hence, we conclude that the flux of XMM0042 and its variability can be ascribed to source 75, with a negligible contribution from source 78.

\begin{deluxetable*}{ccccccccc}
\tablecaption{Best-fit parameters of the total PN, MOS1 and MOS2 spectra of XMM0042 \label{tab:spec}}
\tablecolumns{9}
\tablenum{1}
\tablewidth{0pt}
\tablehead{
Spectral model & nhp & T$_{in}$ & R$_{in}$ & $\Gamma$ & N$_{pow}$ & Line$_E$ & Line$_{\sigma}$ & Line$_{norm}$ \\
& & (keV) & (km) & & 10$^{-5}$ & (keV) & (keV) & 10$^{-5}$\\}
\startdata
{\tt tbabs}({\tt diskbb}+{\tt pow}) & 1.50$\times10^{-10}$ & 0.188$\pm$0.003 & 26.5$\pm1.0$ & 0.98$\pm$0.01 & 6.0$\pm$0.1 & - & - & - \\
{\tt tbabs}({\tt gau}+{\tt diskbb}+{\tt pow}) & 4.11$\times10^{-2}$ & 0.198$\pm$0.003 & 25.5$_{-1.2}^{+1.3}$ & 0.91$\pm$0.01 & 5.5$\pm$0.1 & 0.62$_{-0.03}^{+0.02}$ & 0.15$_{-0.01}^{+0.02}$ & -3.6$_{-0.9}^{+0.6}$\\
{\tt tbabs}({\tt gau}+{\tt diskbb}+{\tt pow}) & 9.49$\times10^{-3}$ & 0.128$\pm$0.005 & 10.5$_{-1.1}^{+1.0}$ & 0.98$\pm$0.01 & 6.0$\pm$0.1 & 0.94$_{-0.03}^{+0.02}$ & 0.20$\pm0.02$ & 1.8$_{-0.3}^{+0.4}$ \\
\enddata
\tablecomments{We report the parameters of best fits obtained using the total PN, MOS1, MOS2 spectra, as discussed in Section \ref{sec:analysis}.
  Inner radius has been calculated using a distance of 784 kpc and an inclination $i$ of 70$^{\circ}$,
  where $R_{in}=D_{[10kpc]}/\sqrt{N_{diskbb}*sin(i)}$
}
\end{deluxetable*}

\begin{deluxetable*}{cccccccc}
\tablecaption{Dipping light curve parameters \label{tab:lc}}
\tablecolumns{10}
\tablenum{2}
\tablewidth{0pt}
\tablehead{
Obs.Num. & OBSID & N$_{dips}$ & nhp & L$_{out}$ & T$_{min}$ & $\Delta$T & L$_{min}$\\
 & & & & $10^{38}$erg s$^{-1}$ & MJD & ks & $10^{38}$erg s$^{-1}$ \\}
\startdata
9 & 0405320701 & 1 & 6.47$\times10^{-2}$ & 0.85$\pm$0.03 & 54100.737$\pm$0.001  & 2.87$\pm$0.47 & 0.17$\pm$0.09\\
11 & 0405320901 & 1 & 3.20$\times10^{-1}$ & 0.93$\pm$0.03 & 54136.287$\pm$0.001 & 3.84$\pm$0.48 & 0.26$\pm$0.06\\
12 & 0505720201 & 2 & 8.56$\times10^{-1}$ & 0.82$\pm$0.02 & 54463.683$\pm$0.001  & 3.04$\pm$0.33 & 0.08$\pm$0.06\\
- & -          & 2 & -                  & -             & 54463.850$\pm$0.002 & 3.36$\pm$0.55 & 0.41$\pm$0.06\\
18 & 0551690201 & 2 & 4.52$\times10^{-3}$ & 0.86$\pm$0.03 & 54830.223$\pm$0.002 & 4.75$\pm$0.64 & 0.16$\pm$0.07\\
- & -          & 2 & -                  & -             & 54830.379$\pm$0.004 & 2.80$\pm$1.76 & 0.22$\pm$0.18\\
22 & 0551690601 & 2 & 8.80$\times10^{-3}$ & 0.82$\pm$0.03 & 54866.623$\pm$0.002 & 2.19$\pm$0.55 & 0.08$\pm$0.02\\
- & -          & 2 & -                  & -             & 54866.791$\pm$0.002 & 2.50$\pm$0.82 & 0.38$\pm$0.11\\
27 & 0600660601 & 1 & 3.14$\times10^{-1}$ & 0.97$\pm$0.03 & 55229.226$\pm$0.002 & 2.23$\pm$0.47 & 0.49$\pm$0.10\\
30 & 0650560401 & 2 & 1.51$\times10^{-2}$ & 0.92$\pm$0.03 & 55576.019$\pm$0.002 & 3.36$\pm$0.55 & 0.21$\pm$0.09\\
- & -          & 2 & -                  & -             & 55576.186$\pm$0.001 & 5.00$\pm$0.56 & 0.10$\pm$0.06\\
33 & 0674210201 & 1 & 9.60$\times10^{-2}$ & 1.38$\pm$0.03 & 55923.159$\pm$0.004 & 5.00$\pm$1.54 & 0.99$\pm$0.11\\
35 & 0674210401 & 1 & 3.61$\times10^{-1}$ & 1.36$\pm$0.03 & 55941.708$\pm$0.001 & 2.83$\pm$0.52 & 0.66$\pm$0.12\\
\enddata
\tablecomments{We report the parameters of the best-fit constant+dips model of dipping {\it XMM-Newton} light curves of XMM0042 we analyzed.
  We show the observation number, number of dips, null hypothesis probability and the parameters as described in Section \ref{sec:analysis}.
  More detailed information can be found in Appendix (Table \ref{tab:lc3b}).}
\end{deluxetable*}

\section{Discussion} \label{sec:disc}

We have discovered a diplike modulation in the light curve of XMM0042. These dips occur with a period of 4.01 h.
If the 4.01 h modulation represents the binary period of XMM0042, the binary separation is $a\sim10^{11}M_X^{1/3}(1+q)^{1/3}$ cm, where
$M_X$ is the mass of the compact object (in solar masses) and $q$ is the mass ratio of the companion star and the compact object.
This short orbital separation rules out a High-Mass X-ray Binary system with a blue supergiant (or a Be main sequence) companion.
This agrees with {\it HST} observations \citep{yuk17} that exclude high-mass ($>3M_{\odot}$) donors at the location of this source.
We note that this period does not exclude an XRB with a Wolf-Rayet donor (Cyg X-3 in our Galaxy has a similar orbital period).
However, the light curve of Cyg X-3 is very different from the one of XMM0042 (very stable and quasi-sinusoidal), so we can safely consider it is more likely a
dipping Low-Mass X-ray Binary (LMXB), given the similarities of their orbital light curves.

LMXB systems are known to show dips if the system is viewed relatively
close to edge-on, i.e. at a high inclination angle of 60--80$^{\circ}$ \citep{fra87}:
the duration and variability of XMM0042 dips is similar to those seen in some well-known Galactic LMXB, such as XB 1254-690 and XB 1916-053 \citep{dia06}.
For these Galactic sources the dips have different shapes and are not detected in all the orbital cycles, as we observe for XMM0042.
Dips are thought to be due to absorption from the matter in the external region of the accretion disk \citep{whi82}.
The spectral evolution during dips depends on the ionization state of the absorbing material \citep{dia16}.
In the case of XMM0042, we do not detect any significant spectral variation possibly due to the low statistics.

\citet{van94} study the optical emission from LMXB disks; if X-rays from the central source are reprocessed by the accretion disk,
this implies that the optical luminosity in the V-band $M_V$ scales with the X-ray luminosity and size of the accretion disk.
We considered the relation from \citet{van94} between $L_X$, the orbital period and $M_V$, taking into account $L_{Edd}=2.5\times10^{38}$ erg s$^{-1}$ \citep[following ][]{van94},
our orbital period of 4.01 hr and our mean luminosity $L_X=2\times10^{38}$ erg s$^{-1}$:

\begin{eqnarray}
  &M_V = 1.57(\pm0.24)-2.27(\pm0.32)log((P/1hr)^{2/3}(L_X/L_{edd})^{1/2}) \nonumber\\
\end{eqnarray}

obtaining $M_V=0.8\pm0.3$ (1$\sigma$ error) for the optical counterpart.
Among the 17 positionally-consistent HST candidate optical/UV counterparts reported in \citet{yuk17}, all but one are very similar to the surrounding stellar populations.
The remaining source shows an excess in the blue band, making it an interesting potential counterpart of the accretion disk of a LMXB.
From the blue apparent magnitude of the counterpart $m_B$=24.78, standard extinction towards M31 and equation 4 from \citet{bar12}, we obtain an absolute visual magnitude $M_V\simeq0.3$.
This is consistent (within 2$\sigma$) with the one derived from the X-ray luminosity and period of XMM0042, thus making it the likely optical counterpart.

The X-ray spectrum of XMM0042 \citep[as seen by][and in this work]{yuk17} is quite hard, with $\Gamma\sim$1 and $E_{cutoff}\sim$18 keV.
This is consistent with what seen in dipping LMXBs in our Galaxy, where a wide range of spectral properties is diplayed,
with photon indexes varying from 0.4 to 2 and cutoff energies from 3.5 to 80 keV \citep{dia06}.
The most luminous Galactic dipper, X 1624-490, has an X-ray luminosity in the 1-30 keV energy band of $\sim7.3\times10^{37}$ erg s$^{-1}$
\citep[in the 0.5-50 keV energy range -- Galactic dippers reach persistent luminosities only $\sim3$ times lower than XMM0042,][]{bal00,iar07}
and a much softer spectrum.

The period, optical counterpart and the spectrum of XMM0042 are reminiscent of a dipping LMXB system.
The mean luminosity we found through our {\it XMM-Newton} analysis, ranging from 0.8 to 2.8$\times10^{38}$ erg s$^{-1}$, makes XMM0042 the most luminous dipper to date.
\citet{tru02} found the first dipping source of M31, located in Bo 158.
The period of the dipping behavior was 2.78 hr with a flux modulation of $\sim$83\%. A study of the other {\it XMM-Newton}, {\it Chandra}
and {\it ROSAT} observations \citep{tru04} revealed the amplitude of the dips to be anti-correlated with the source luminosity (0.5 to 2$\times 10^{38}$ erg s$^{-1}$),
disappearing at high luminosities. The source also showed hour-timescale and months-timescale luminosity variations.
This source has a rather hard spectrum but it never been observed by {\it NuSTAR}, so that we have no information about the cutoff
energy or its luminosity in the hard X-ray band. Another possibly dipping source in M31 was identified by \citet{man04}, with a period of
$\sim$1.8 hr and a 0.3-10 keV luminosity of $\sim10^{37}$ erg s$^{-1}$. In this case, however, a foreground X-ray source cannot be ruled out.
We argue that we are observing the most luminous dippers in M31, the bright tail of their population in M31.

\acknowledgments

This research has made use of data produced by the EXTraS project, funded by the European Union's Seventh Framework Programme under grant agreement no 607452.
The EXTraS project acknowledges the usage of computing facilities at INAF’s Astronomical Observatory of Catania.
The EXTraS project acknowledges the CINECA award under the ISCRA initiative, for the availability of high performance computing resources and support.

\vspace{5mm}
\facilities{XMM}

\clearpage

\appendix

\startlongtable
\begin{deluxetable*}{cccc}
\tablecaption{Parameters for all the analyzed observations \label{tab:lc3a}}
\tablehead{
Obs.Num. & OBSID & Obs.date & Livetime\\
 & & & ks\\}
\startdata
1 & 0112570401 & 2000-06-25T10:44:42 & 34.5 \\
2 & 0109270101 & 2001-06-29T06:59:13 & 54.4 \\
3 & 0112570101 & 2002-01-06T18:07:17 & 63.0 \\
4 & 0202230201 & 2004-07-16T16:17:49 & 19.7 \\
5 & 0202230401 & 2004-07-18T23:50:10 & 21.4 \\
6 & 0202230501 & 2004-07-19T12:49:04 & 26.1 \\
7 & 0405320501 & 2006-07-02T14:14:29 & 21.3 \\
8 & 0405320601 & 2006-08-09T11:59:23 & 21.3 \\
9 & 0405320701 & 2006-12-31T14:01:28 & 15.4 \\
10 & 0405320801 & 2007-01-16T11:24:02 & 13.4 \\
11 & 0405320901 & 2007-02-05T03:21:01 & 16.4 \\
12 & 0505720201 & 2007-12-29T13:19:11 & 26.8 \\
13 & 0505720301 & 2008-01-08T06:38:03 & 26.5 \\
14 & 0505720401 & 2008-01-18T14:48:44 & 22.2 \\
15 & 0505720501 & 2008-01-27T22:05:19 & 21.2 \\
16 & 0505720601 & 2008-02-07T04:33:16 & 21.3 \\
17 & 0560180101 & 2008-07-18T05:49:32 & 21.3 \\
18 & 0551690201 & 2008-12-30T03:04:26 & 21.2 \\
19 & 0551690301 & 2009-01-09T05:56:27 & 21.1 \\
20 & 0551690401 & 2009-01-15T21:17:25 & 26.3 \\
21 & 0551690501 & 2009-01-27T06:59:35 & 21.0 \\
22 & 0551690601 & 2009-02-04T12:57:36 & 19.5 \\
23 & 0600660201 & 2009-12-28T12:19:26 & 18.1 \\
24 & 0600660301 & 2010-01-07T07:23:07 & 16.7 \\
25 & 0600660401 & 2010-01-15T12:20:26 & 16.6 \\
26 & 0600660501 & 2010-01-25T02:15:46 & 19.1 \\
27 & 0600660601 & 2010-02-02T02:18:08 & 16.8 \\
28 & 0650560201 & 2010-12-26T09:55:44 & 26.3 \\
29 & 0650560301 & 2011-01-04T17:46:48 & 32.7 \\
30 & 0650560401 & 2011-01-14T23:53:29 & 23.7 \\
31 & 0650560501 & 2011-01-25T06:52:18 & 23.3 \\
32 & 0650560601 & 2011-02-03T23:34:45 & 23.3 \\
33 & 0674210201 & 2011-12-28T00:45:21 & 20.3 \\
34 & 0674210301 & 2012-01-07T02:24:46 & 16.8 \\
35 & 0674210401 & 2012-01-15T14:37:17 & 19.4 \\
36 & 0674210501 & 2012-01-21T11:58:42 & 16.8 \\
37 & 0674210601 & 2012-01-31T01:56:08 & 20.9 \\
38 & 0700380501 & 2012-07-28T14:53:08 & 11.5 \\
39 & 0700380601 & 2012-08-08T22:44:48 & 23.4 \\
40 & 0727960401 & 2013-07-06T07:43:22 & 10.5 \\
\enddata
\tablecomments{We show the observation number and characteristics of each {\it XMM-Newton} observation we analyzed.}
\end{deluxetable*}

\startlongtable
\begin{deluxetable*}{ccccccccc}
\tablecaption{Light curves Parameters for all the analyzed observations \label{tab:lc3b}}
\tablehead{
Obs.Num. & Best fit & DoF & nhp & L$_{out}$ & lin & T$_{min}$ & $\Delta$T & L$_{min}$\\
 & & & & $10^{38}$erg s$^{-1}$ & & MJD & ks & $10^{38}$erg s$^{-1}$ \\}
\startdata
1 & constant & 70 & 8.26$\times10^{-2}$ & 1.13$\pm$0.02 & - & - & - & -\\
2 & constant & 110 & 2.15$\times10^{-1}$ & 1.16$\pm$0.01 & - & - & - & -\\
3 & linear & 126 & 1.42$\times10^{-1}$ & 1.04$\pm$0.03 & (-5.1$\pm$0.7)$\times10^{-6}$ & - & - & -\\
4 & constant & 40 & 1.78$\times10^{-1}$ & 1.38$\pm$0.03 & - & - & - & -\\
5 & constant & 38 & 5.98$\times10^{-1}$ & 1.09$\pm$0.02 & - & - & - & -\\
6 & constant & 53 & 6.14$\times10^{-1}$ & 0.93$\pm$0.02 & - & - & - & -\\
7 & constant & 43 & 7.86$\times10^{-2}$ & 1.68$\pm$0.03 & - & - & - & -\\
8 & constant & 43 & 9.96$\times10^{-1}$ & 1.21$\pm$0.02 & - & - & - & -\\
9 & con+1dip & 28 & 6.47$\times10^{-2}$ & 0.85$\pm$0.03 & - & 54100.737$\pm$0.001 & 2.87$\pm$0.47 & 0.17$\pm$0.09\\
10 & constant & 28 & 7.30$\times10^{-1}$ & 1.56$\pm$0.03 & - & - & - & -\\
11 & con+1dip & 30 & 3.20$\times10^{-1}$ & 0.93$\pm$0.03 & - & 54136.287$\pm$0.001 & 3.84$\pm$0.48 & 0.26$\pm$0.06\\
12 & con+2dip & 48 & 8.56$\times10^{-1}$ & 0.82$\pm$0.02 & - & 54463.683$\pm$0.001 & 3.04$\pm$0.33 & 0.08$\pm$0.06\\
- & con+2dip & - & - & - & - & 54463.850$\pm$0.002 & 3.36$\pm$0.55 & 0.41$\pm$0.06\\
13 & constant & 54 & 3.63$\times10^{-2}$ & 1.66$\pm$0.03 & - & - & - & -\\
14 & constant & 45 & 4.03$\times10^{-2}$ & 1.49$\pm$0.03 & - & - & - & -\\
15 & constant & 43 & 3.61$\times10^{-1}$ & 1.15$\pm$0.02 & - & - & - & -\\
16 & constant & 44 & 3.53$\times10^{-1}$ & 1.61$\pm$0.02 & - & - & - & -\\
17 & constant & 44 & 2.83$\times10^{-1}$ & 1.45$\pm$0.04 & - & - & - & -\\
18 & con+2dip & 37 & 4.52$\times10^{-3}$ & 0.86$\pm$0.03 & - & 54830.223$\pm$0.002 & 4.75$\pm$0.64 & 0.16$\pm$0.07\\
- & con+2dip & - & - & - & - & 54830.379$\pm$0.004 & 2.80$\pm$1.76 & 0.22$\pm$0.18\\
19 & constant & 44 & 5.92$\times10^{-1}$ & 1.51$\pm$0.02 & - & - & - & -\\
20 & constant & 53 & 2.49$\times10^{-1}$ & 1.23$\pm$0.03 & - & - & - & -\\
21 & constant & 43 & 8.64$\times10^{-2}$ & 1.47$\pm$0.03 & - & - & - & -\\
22 & con+2dip & 48 & 8.80$\times10^{-3}$ & 0.82$\pm$0.03 & - & 54866.623$\pm$0.002 & 2.19$\pm$0.55 & 0.08$\pm$0.02\\
- & con+2dip & - & - & - & - & 54866.791$\pm$0.002 & 2.50$\pm$0.82 & 0.38$\pm$0.11\\
23 & constant & 37 & 2.13$\times10^{-1}$ & 1.37$\pm$0.03 & - & - & - & -\\
24 & constant & 34 & 6.84$\times10^{-1}$ & 1.02$\pm$0.02 & - & - & - & -\\
25 & constant & 34 & 3.83$\times10^{-1}$ & 1.28$\pm$0.03 & - & - & - & -\\
26 & constant & 39 & 9.41$\times10^{-2}$ & 1.34$\pm$0.03 & - & - & - & -\\
27 & con+1dip & 31 & 3.14$\times10^{-1}$ & 0.97$\pm$0.03 & - & 55229.226$\pm$0.002 & 2.23$\pm$0.47 & 0.49$\pm$0.10\\
28 & constant & 53 & 1.66$\times10^{-2}$ & 1.13$\pm$0.02 & & - & - & - \\
29 & constant & 66 & 4.84$\times10^{-1}$ & 1.69$\pm$0.02 & - & - & - & -\\
30 & con+2dip & 42 & 1.51$\times10^{-2}$ & 0.92$\pm$0.03 & - & 55576.019$\pm$0.002 & 3.36$\pm$0.55 & 0.21$\pm$0.09\\
- & con+2dip & - & - & - & - & 55576.186$\pm$0.001 & 5.00$\pm$0.56 & 0.10$\pm$0.06\\
31 & constant & 47 & 1.42$\times10^{-1}$ & 1.47$\pm$0.03 & - & - & - & -\\
32 & constant & 47 & 1.97$\times10^{-1}$ & 1.69$\pm$0.03 & - & - & - & -\\
33 & con+1dip & 38 & 9.60$\times10^{-2}$ & 1.38$\pm$0.03 & - & 55923.159$\pm$0.004 & 5.00$\pm$1.54 & 0.99$\pm$0.11\\
34 & constant & 34 & 8.59$\times10^{-1}$ & 2.54$\pm$0.03 & - & - & - & -\\
35 & con+1dip & 36 & 3.61$\times10^{-1}$ & 1.36$\pm$0.03 & - & 55941.708$\pm$0.001 & 2.83$\pm$0.52 & 0.66$\pm$0.12\\
36 & variable & 34 & 2.63$\times10^{-5}$ & 1.47$\pm$0.05 & - & - & - & -\\
37 & constant & 46 & 7.00$\times10^{-2}$ & 2.83$\pm$0.04 & - & - & - & -\\
38 & constant & 23 & 3.99$\times10^{-1}$ & 1.79$\pm$0.09 & - & - & - & -\\
39 & constant & 47 & 1.67$\times10^{-1}$ & 2.66$\pm$0.07 & - & - & - & -\\
40 & constant & 21 & 8.35$\times10^{-1}$ & 2.18$\pm$0.04 & - & - & - & -\\
\enddata
\tablecomments{We report the parameters of the best-fit model of each {\it XMM-Newton} light curve we analyzed.
  We show the observation number, best fitting model, degrees of freedom, null hypothesis probability and the parameters as described in the paper.
  The curves are in the 0.2-12 keV energy range, produced using EXTraS tools.
  We used spectral information to convert count rate into luminosity and then to united the curve of each exposure and camera within the same observation.
  For the linear model, the constant parameter is evaluated at half observation.}
\end{deluxetable*}

\end{document}